\documentclass[12pt,a4paper]{conference}

\usepackage{psfrag}
\usepackage{epsfig}
\usepackage{fancyhdr}
\usepackage{graphicx,amsmath,amssymb,cite}
\usepackage{multind}
\makeindex{author} \makeindex{subject}
\newcommand{\beq}{\begin{equation}}
\newcommand{\eeq}[1]{\label{#1}\end{equation}}
\newcommand{\eeqn}{\end{equation}}

\newcommand{\beqa}{\begin{eqnarray}}
\newcommand{\eeqa}[1]{\label{#1}\end{eqnarray}}
\newcommand{\eeqan}{\end{eqnarray}}

\let\bar=\overbar

\newcommand{\Dslash}{\not{\hbox{\kern-4pt $D$}}}
\newcommand{\dslash}{\not{\hbox{\kern-2pt $\del$}}}

\newcommand{\msb}{{\bar{\ssstyle M \kern -1pt S}}}

\begin{document}

\Chapter{Progress in $NN\to NN\pi$}
           {Progress in $NN\to NN\pi$}{V. Baru \it{et al.}}
\vspace{-5 cm}\includegraphics[width=6 cm]{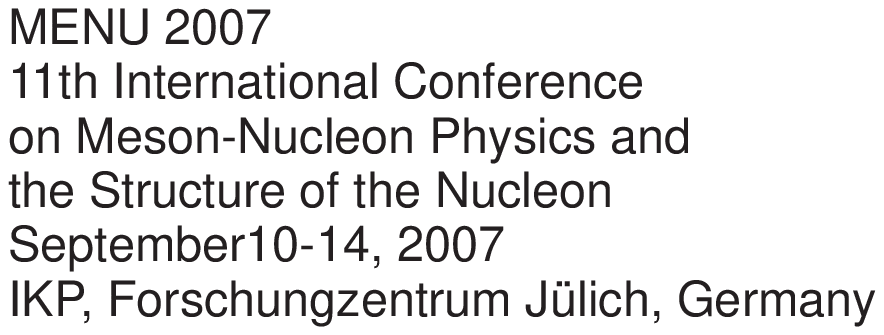}
\vspace{4 cm}

\addcontentsline{toc}{chapter}{{\it N. Author}} \label{authorStart}

\begin{raggedright}

\underline{V. Baru}$^{\star}$$^,$\footnote{E-mail address: baru@itep.ru}~, 
J. Haidenbauer$^{\%}$,  C.~Hanhart$^{\%}$,
A. Kudryavtsev$^{\star}$, V. Lensky$^{\star,\%}$ and U.-G. Mei\ss ner$^{\%,\#}$

\bigskip\bigskip


$^{\star}$Institute of Theoretical and Experimental Physics,
 117259, B.~Cheremushkinskaya 25, Moscow, Russia \\
$^{\%}$Institut f\"{u}r Kernphysik, Forschungszentrum J\"{u}lich GmbH,
 D--52425 J\"{u}lich, Germany\\
$^{\#}$Helmholtz-Institut f\"{u}r Strahlen- und Kernphysik (Theorie), 
Universit\"at Bonn,  Nu{\ss}allee 14-16, D--53115 Bonn, Germany \\

\end{raggedright}


\begin{center}
\textbf{Abstract}
\end{center}

We survey the recent developments in the reaction $NN \to NN\pi$  in
 effective field theory. 
We show that the proper construction of the production operator needs a 
careful separation of irreducible pieces from reducible ones. The result of 
this consideration is a complete cancellation of all loops in the production 
operator at NLO. Moreover, we show that this procedure brings the leading 
Weinberg-Tomozawa vertex on-shell, thus enhancing the corresponding 
contribution to the transition amplitude by a factor of 4/3 as compared to 
the commonly used value.
We also discuss the role of the $\Delta(1232)$ for the $s$-wave pion production.
Being relatively sizable individually the direct and rescattering mechanisms of the $\Delta$ 
excitation at NLO cancel each other to a large extent. Thus, we conclude that the net effect of the 
$\Delta$ at NLO is very small. 

\section{Introduction}

Understanding the dynamics of pion production in nucleon-nucleon collisions near
threshold is a challenge for theoreticians. 
Knowledge of the pion production mechanism in the isospin symmetric case 
is an important step to the study of isospin violation in few-nucleon 
processes\cite{vKMiNi00,Gaa04}, which provides a test for chiral 
perturbation theory (ChPT). Furthermore, success in the description of the 
charged pion production reactions is a necessary condition for a calculation 
of the dispersive correction to $\pi d$ scattering\cite{VL06_1}, which is one of the
most uncertain and at the same time important corrections to this process.
When accurate data for the total cross-section close to threshold appeared 
in 1990\cite{IUCF90}, existing models\cite{KR,Mi90} failed
to describe the data by a factor of five to ten for the channel
$pp\to pp\pi^0$ and a factor of two for the channels $pp\to pn\pi^+$ and
$pp\to d\pi^+$. 
To cure this discrepancy, many phenomenological mechanisms were proposed
 --- for a recent review see Ref.~\cite{CH04}.
Also various groups started to investigate $NN\to NN\pi$ using ChPT. As a big surprise,
 however, it turned out that using the original power counting
 proposed by Weinberg \cite{swein1} leads to even larger discrepancy between
data and theory  
at next--to--leading order (NLO) for
 $pp\to pp\pi^0$ \cite{park} as well as for $pp\to d\pi^+$
 \cite{unserd}. 
 Even worse, the corrections at one--loop order
 (next--to--next--to--leading order (N$^2$LO) in the standard counting) turned
 out to be even larger than the tree level NLO corrections, putting into question the convergence
 of  the chiral expansion \cite{dmitrasinovic,ando}.

 At the same time it was already realized that a
 modified power counting is necessary to properly take care of the large
 momentum transfer characteristic for pion production in $NN$ collisions 
\cite{Co96,dR00,CH00,HanKai}.
The expansion parameter in this case is 
\begin{equation}
\chi=\frac{|\vec p_{\mathrm{thr}}|}{M_N}=\sqrt{\frac{m_\pi}{M_N}}
\end{equation}
 where $m_\pi\, (M_N)$ is the pion (nucleon) mass and $|\vec p_{\mathrm{thr}}|$ is 
the initial nucleon momentum at threshold. 
As a consequence the hierarchy of diagrams changes and some one-loop diagrams 
start to contribute already at NLO.  
In this presentation we discuss the charged pion production  where the produced pion is in an s-wave relative to the final NN pair up to NLO in ChPT.
In sec. 2 we discuss the pion production operators involving only 
pionic and nucleonic degrees of freedom. 
We start from the concept of reducibility  that is necessary
to distinguish between the production operator, which should consist of all irreducible pieces, and the NN wave functions. The proper treatment of this concept allows to avoid double counting in 
the calculation and thus is extremely important. We discuss in detail the special case of diagrams
with the energy dependent vertices originating from, e.g., the Weinberg-Tomozawa (WT) term. 
In this case some diagrams that seem to be purely reducible from the general rules 
acquire an important irreducible contribution. 
In sec. 3 we discuss the role of the Delta resonance for the s-wave pion production near threshold. 
The corresponding diagrams start to contribute at NLO and thus are relevant for 
the present study.  The main results are summarized in sec. 4.

\section{Nucleonic amplitudes up to NLO}

\begin{figure}[t]
\begin{center}
\epsfig{file=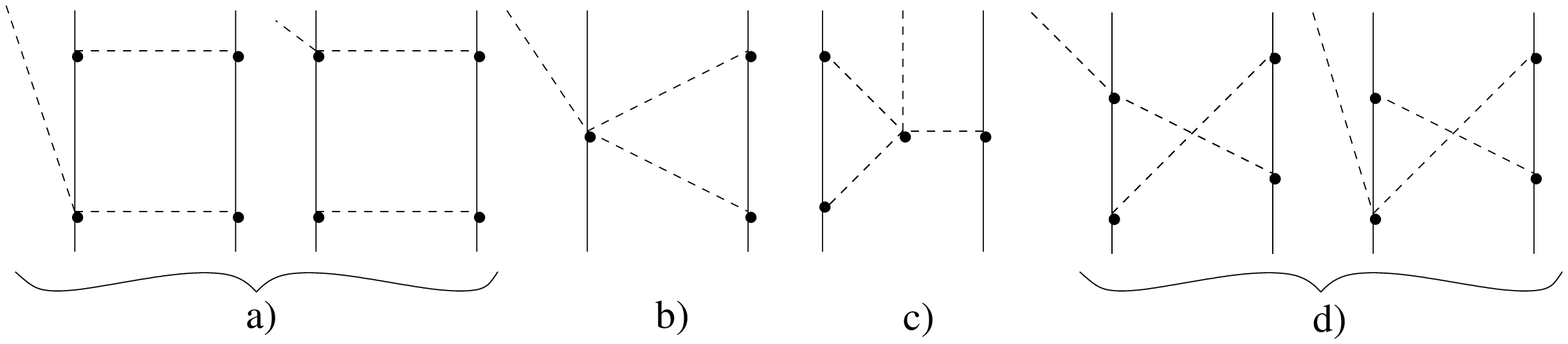, height=2.5cm, angle=0}
\caption{Leading loop diagrams for $NN\to NN\pi$. Here dashed lines denote
  pions and solid lines denote nucleons.}
\label{diagram}
\end{center}
\end{figure}

A method  how to calculate processes on few
nucleon systems with external probes was proposed by Weinberg \cite{swein1}: here the transition 
(production) operators are to be calculated using ChPT. Then those
transition operators must be convoluted with the appropriate $NN$ wave
functions --- in full analogy to the so--called distorted wave Born
approximation traditionally used in phenomenological calculations~\cite{KR}.

Therefore it is necessary to disentangle those diagrams that are part of the
wave function from those that are part of the transition operator. In complete
analogy to $NN$ scattering, the former are called reducible
and the latter irreducible. The distinction stems from whether or
not the diagram shows a two-nucleon cut. Thus, in accordance to this rule,
 the one-loop diagrams shown in
Fig. \ref{diagram}(b)--(d) are irreducible, whereas diagrams (a)  seem to be
reducible. This logic was used in the paper by Hanhart and Kaiser \cite{HanKai}  
to  single out the irreducible loops contributing at NLO.
The findings of Ref.~\cite{HanKai} were:
\begin{itemize}
\item For the channel $pp\to pp\pi^0$ the sum of diagrams (b)--(d) of Fig.
  \ref{diagram} vanished due to a cancellation between individual diagrams
\item For the channel $pp\to d\pi^+$ the same sum gave a finite answer\footnote{The
connection of the amplitude $A$ to the observables is given, e.g., in Ref~\cite{Lensky}}:
\begin{equation}
A_{pp\to d\pi^+}^{b+c+d} = \frac{g_A^3}{256f_\pi^5} \left(-2+3+0\right)\,
|\vec q|=\frac{g_A^3|\vec q|}{256f_\pi^5}.
\end{equation}
\end{itemize}\noindent
The latter amplitude grows linearly with increasing final
 $NN$--relative momentum  $|\vec q|$, which leads to a large
  sensitivity to the final $NN$ wave function, once the convolution of those
  with the transition operators is evaluated. However, the problem is that such a
 sensitivity is not allowed in a consistent field theory as was stated
 in Ref.~\cite{Gaa05}. 
The solution of this problem was presented in Ref.~\cite{Lensky} and will be discussed in this presentation. 
 
It is the main point of this section that diagrams
(a) contain a genuine irreducible piece due to the energy dependence of the
leading $\pi N\to \pi N$ vertex.
Specifically, the energy dependent part of the WT vertex cancels one of the 
intermediate nucleon propagators, resulting in the additional irreducible 
contribution at NLO from diagrams (a). This additional contribution compensates the linear 
growth of diagrams (b)--(d) thus solving the problem. 
\begin{figure}[t!]
\begin{center}
\psfrag{xx1}{$(m_\pi,\vec 0)$}
\psfrag{xx2}{$(l_0,\vec l)$}
\psfrag{yy1}{$(E+l_0-m_\pi,\vec p+\vec l)$}
\psfrag{yy2}{$(E,\vec p)$}
\psfrag{VV}{\Large $V_{\pi\pi NN}=$}
\epsfig{file=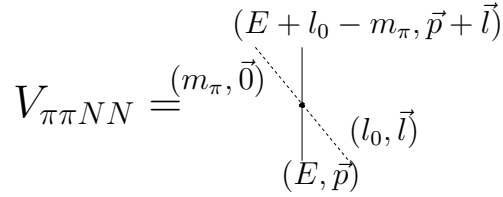, height=2.2cm}
\end{center}
\caption{The $\pi N\to \pi N$ transition
vertex: definition of kinematic variables as used in the text.}
\label{vpipinn}
\end{figure}
To demonstrate this, we write the expression for the $\pi N\to \pi N$ vertex 
in the notation of Fig. 2 as 

\begin{eqnarray}
V_{\pi\pi NN}&=&
l_0{+}m_\pi{-}\frac{\vec l\cdot(2\vec p+\vec l)}{2M_N} \nonumber \\
&=&
\underbrace{{2m_\pi}}_{\mbox{on-shell}}{+}\underbrace{{\left(l_0{-}m_\pi{+}E{-}
\frac{(\vec l+\vec p)^2}{2M_N}\right)}}_{(E'-H_0)=(S')^{-1}}{-}\underbrace{{\left(E{-}\frac{\vec p\,
    ^2}{2M_N}\right)}}_{(E-H_0)=S^{-1}} \ .
\label{pipivert}
\end{eqnarray}
For simplicity we skipped the isospin part of the amplitude. The first term in
the last line denotes the transition in on--shell kinematics, the second the
inverse of the outgoing nucleon propagator and the third the inverse of the
incoming nucleon propagator.  Note that for on--shell
incoming and outgoing nucleons, the $\pi N\to \pi N$ transition vertex
takes its on--shell value $2m_\pi$ --- even if the incoming pion is
off--shell. 
This is in contrast to standard phenomenological treatments~\cite{KR}, where $l_0$ is
identified with $m_\pi/2$ --- the energy transfer in on--shell kinematics ---
and the recoil terms are not considered. Note, since $p_{thr}^2/M_N=m_\pi$
the recoil terms are to be kept.
\begin{figure}[ht!]
\begin{center}
\includegraphics[height=3.3cm,width=12.0cm,keepaspectratio]{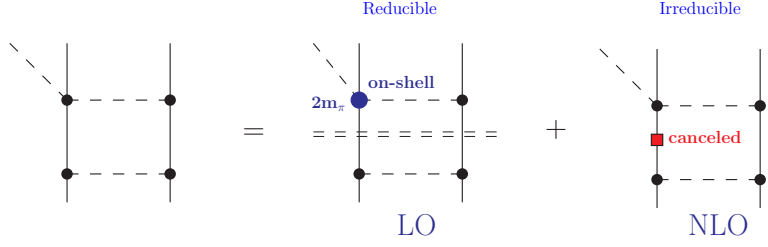}
\caption{Reducible and irreducible parts of the box diagram. The square indicates
that the corresponding nucleon propagator is canceled. The double dashed line shows the two-nucleon cut.}
\label{Pattern}
\end{center}
\end{figure}
A second consequence of Eq. (\ref{pipivert}) is even more interesting: when
the rescattering diagram with the $\pi N\to \pi N$ vertex gets convoluted with $NN$ wave functions, only the
first term leads to a reducible diagram. The second and third terms, however,
lead to irreducible contributions, since one of the nucleon propagators gets
canceled. This is illustrated in Fig. \ref{Pattern} on the example of the second diagram of Fig. \ref{diagram}a.  
It was shown explicitly in Ref. \cite{Lensky} that those induced irreducible contributions
cancel the finite remainder of the NLO loops ((b)-(d)) in the $pp\to d\pi^+$ channel.
\begin{figure}[h!]
\begin{center}
\includegraphics[height=5.5cm,width=14.0cm,keepaspectratio]{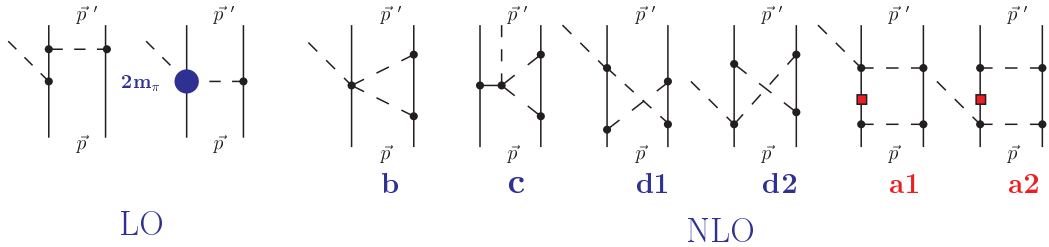}
\caption{Complete set of nucleonic diagrams up to NLO. 
Note that sum of all loops at NLO vanishes.}
\label{NLOdiagr}
\end{center}
\end{figure}
Thus, up to NLO only the diagrams appearing at LO, as shown in Fig. \ref{NLOdiagr}, 
contribute to $pp\to d\pi^+$, with the rule that the $\pi N\to \pi N$ vertex is put 
on--shell. This enhances the dominating isovector $\pi N$-rescattering amplitude by a 
factor of $4/3$ as compared to the traditionally used value, 
which leads to a good description of the experimental data for 
$pp\to d\pi^+$. The result found in Ref. \cite{Lensky} is shown in Fig. \ref{NNpi_result}
as the solid line whereas the dashed line
is the result of the model by Koltun and Reitan~\cite{KR} that basically corresponds to our
LO calculation with $3/2 m_{\pi}$ for the $\pi N \to \pi N$ vertex. 
The data sets are from TRIUMF~\cite{dpidata1}, IUCF~\cite{dpidata2},
COSY~\cite{dpidata3} and PSI~\cite{dgotta}.
\begin{figure}[t!]
\begin{center}
\psfig{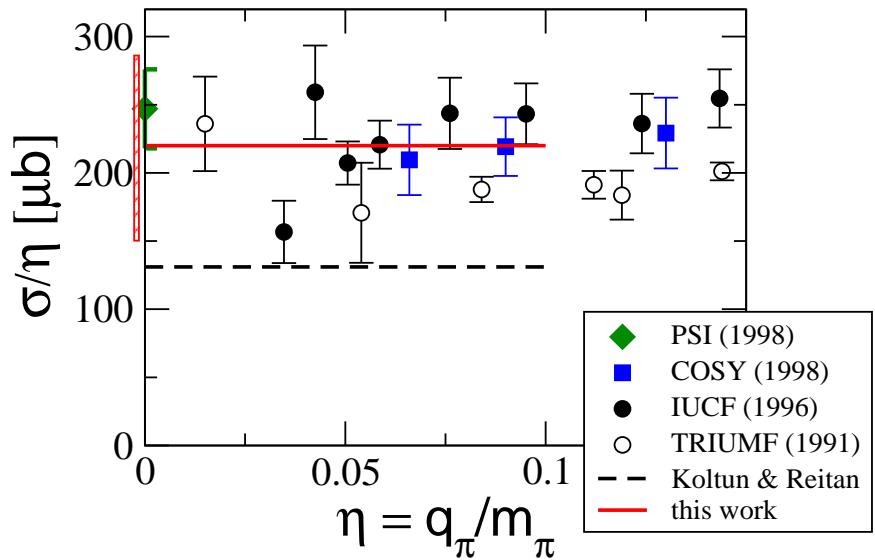}
\end{center}
\caption{Comparison of our results
 to experimental data for $NN\to d\pi$ (see text for the details). 
}
\label{NNpi_result}
\end{figure}

\section{Role of the Delta resonance}

In this section we would like to discuss the influence of the $\Delta$
resonance on the s-wave pion production cross section. The effect of
the Delta has been extensively studied in the literature both in the
phenomenological framework and using EFT. However, the current
situation in the literature is quite contradictory. In particular, in
phenomenological study by Niskanen \cite{Nisk} it was shown that the inclusion of
the $\Delta$ isobar leads to an enhancement of the total cross section in $pp\to
d\pi^+$  by almost a factor of 3.  This enhancement is governed by the
process where $\Delta$ in the intermediate state emits a p-wave pion
which is then rescattered on the nucleon in an s-wave, i.e.
the diagrams  analogous to 
the box diagrams of Fig.\ref{diagram} a) but with the $\Delta$ instead
of the nucleon in the intermediate state. 
However, the finding of Ref.~\cite{Nisk} was not confirmed in the model
calculation by Hanhart et al. \cite{HanJuelich}. The authors of this work have found that
the direct pion emission from the $\Delta$ increases the cross section by about 30\% 
whereas the rescattering process is negligible. 
The J{\"u}lich meson-nucleon model \cite{Schutz} was applied in Ref.\cite{HanJuelich} to
generate the off-shell $\pi N\to \pi N$ transition T-matrix and 
the coupled channel CCF model \cite{ccf} was employed to take into account the $N\Delta$ and $NN$
distortions in the initial and final states. 
In contrary to these results the direct pion production from the $\Delta$ was
shown to be negligible in the EFT calculation by da Rocha et al. \cite{Rocha}.
However, the $NN\to N\Delta$ transition in Ref \cite{Rocha} was
approximated by one-pion exchange only, and the $NN\to N\Delta$ contact interaction that contributes at the same order
was not taken into account. At the same time it is known from
phenomenology that heavy $\rho$ meson exchange which plays the role of this contact term
in phenomenological calculations is significant \cite{ccf}. 
In addition the static $\Delta$ propagator is used in Ref.~\cite{Rocha}
which leads to the large model dependence of the results. 
The similar problem with the use of the static $\Delta$ propagator for
the $\pi d$ scattering was investigated in Ref.~\cite{Baru}.
\begin{figure}[t!]
\begin{center}
\includegraphics[height=3.5cm,width=13.0cm,keepaspectratio]{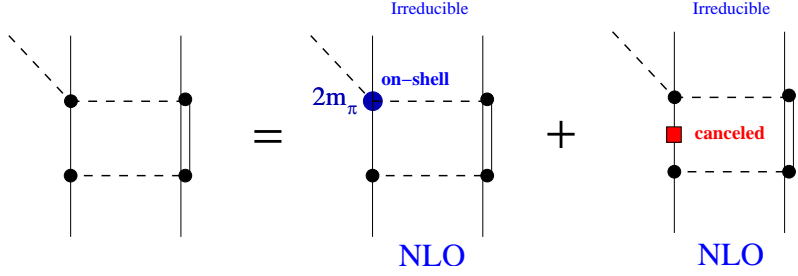}
\end{center}
\caption{Rescattering diagram with the $\Delta$ isobar. Pattern of
  separation on two parts due to the energy dependent WT vertex.  }
\label{PatternDelta}
\end{figure}
Let us now discuss in more detail the $\pi N$ rescattering diagrams
with the $\Delta$. First of all, due to the small mass difference
between the nucleon and $\Delta$, $\Delta M= M_{\Delta}-M_N\simeq 2m_{\pi}$, the
$N\Delta$ propagator  behaves as  
\begin{equation} 
\frac{1}{m_{\pi}-\Delta M - p^2/2M_{N\Delta}} \sim
  \frac{1}{m_{\pi}},
\end{equation}
where $M_{N\Delta}$ is the reduced mass of the $N\Delta$ system, i.e. 
in full analogy with the counting rules for the NN
propagator. Secondly, these rescattering diagrams contain the energy
dependent WT vertex, and thus the method developed in the previous
section for the diagrams with nucleons can be applied here as well. 
In particular, these diagrams can be also divided into two parts, as
demonstrated in Fig. \ref{PatternDelta}:  the first one goes with the on
shell $\pi N$ vertex and in the second one the nucleon propagator
cancels  the corresponding piece in the vertex. In
contradistinction from the diagrams of Fig. \ref{Pattern} both these
parts contribute at NLO and both are irreducible.   Furthermore, in 
the recent paper by Hanhart and Kaiser \cite{HanKai} the full set
of one-loop diagrams with the $\Delta$ contributing at NLO was studied
in EFT. In particular, it was shown that those parts of the rescattering
diagrams, in which the nucleon propagator is canceled, 
take part in a cancellation with other loop diagrams (see
Fig.~4 in Ref. \cite{HanKai}). Thus, the only remainders
contributing at NLO are the direct pion production  
and the rescattering process with the on shell $\pi N$ vertex as shown in
Fig.~\ref{DeldiagNLO}.
\begin{figure}[t!]
\begin{center}
\includegraphics[height=2.3cm,width=12.0cm,keepaspectratio]{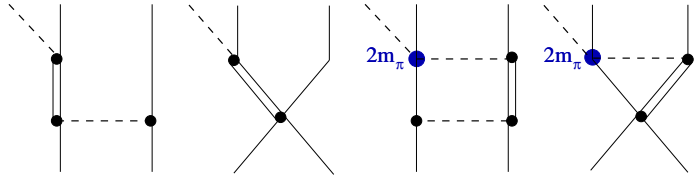}
\end{center}
\caption{Non vanishing diagrams with the $\Delta$ isobar contributing at NLO.}
\label{DeldiagNLO}
\end{figure}
We evaluated these diagrams in a manner similar to our recent study of the role of the $\Delta$
in pion-deuteron scattering \cite{Baru}. 
The calculation revealed that each of these diagrams gives about a 10-15\% correction 
to the transition amplitude but they enter with opposite signs -- 
the direct contribution 
increases the cross section in line with the finding of Ref. \cite{HanJuelich}
whereas the rescattering piece leads to a reduction by almost the same amount. 
Thus, there is a significant cancellation between different mechanisms involving the $\Delta$ excitation and the resulting contribution of the $\Delta$ resonance at NLO is very small.
The calculation was done with the CCF \cite{ccf} and the Hannover \cite{hannover} 
coupled-channel NN models, and the 
pattern of cancellation was the same for both models although the individual contributions 
were slightly different. 

\section{Summary}

We reported about recent developments in the reaction $NN \to NN\pi$  in
the effective field theory.  Within the counting scheme that
acknowledges the large momentum transfer between the initial and the final
nucleons  we have calculated the transition operator for this reaction up to NLO. 
We discussed how to implement properly the reducibility concept for 
the pion production process. In particular it was shown how to identify the 
irreducible contribution of the formally reducible loop diagrams with the
energy dependent vertices. As a result we obtained that the irreducible 
loops at NLO cancel altogether, and the leading order $\pi N$-rescattering amplitude 
is enhanced by a factor of $4/3$ as compared to the commonly used value. This enhancement  
leads to a good description of the experimental data for $pp\to d\pi^+$. 
We also investigated the effect of the $\Delta$ isobar on the s-wave 
pion production. Being relatively sizable individually the direct and rescattering mechanisms of 
the $\Delta$ excitation at NLO cancel each other to a large extent. 
Thus, we conclude that the net effect of the $\Delta$ at NLO is very small. 

The theoretical uncertainty of our NLO calculation was estimated conservatively using the dimensional arguments. 
The large uncertainty of about $2 m_{\pi}/M_N\approx 30\%$ for the cross section is a 
consequence of the rather large expansion parameter.
Thus, a computation at NNLO is necessary for drawing more solid conclusions on the pion production 
mechanism, especially if one wants to learn more about the charge-symmetry breaking effects.

\vspace*{-0.5cm}
\section*{Acknowledgments}
This research is part of the EU Integrated Infrastructure Initiative
Hadron Physics Project under contract number RII3-CT-2004-506078, and
was supported also by the DFG-RFBR grant no. 05-02-04012 (436 RUS
113/820/0-1(R)) and the DFG SFB/TR 16 "Subnuclear Structure of Matter".  V.~B and
A.~K. acknowledge the support of the Federal Agency of Atomic Research
of the Russian Federation.

\end{document}